\documentclass[12pt,english,reprint,aps,prb,superscriptaddress]{revtex4-2}
\usepackage[T1]{fontenc}
\usepackage[utf8]{inputenc}
\usepackage{amstext}
\usepackage{amssymb}
\usepackage{graphicx}
\usepackage{adjustbox}
\usepackage{tabularx}
\usepackage{amsmath}
\makeatletter
\usepackage{babel}
\usepackage[unicode=true]{hyperref}
\usepackage{newtxtext}
\usepackage[slantedGreek,varg,varbb]{newtxmath}
\usepackage{mathtools}
\usepackage{siunitx}
\makeatother
\usepackage{xcolor}   

\usepackage[normalem]{ulem} 
\begin{document}

% TITLE
\title {Signatures of hydrodynamic flow of topological carriers in SnTe multi-terminal nanowires }

\author{Dawid~Śnieżek}
\affiliation{Institute of Physics, Polish Academy of Sciences, Aleja Lotników 32/46, PL-02-668 Warsaw, Poland}

\author{Cezary~Śliwa}
\affiliation{Institute of Physics, Polish Academy of Sciences, Aleja Lotników 32/46, PL-02-668 Warsaw, Poland}
\affiliation{International Research Centre MagTop, Institute of Physics. Polish Academy of Sciences, Aleja Lotnikow 32/46, PL-02-668 Warsaw, Poland}

\author{Krzysztof Dybko}
\affiliation{Institute of Physics, Polish Academy of Sciences, Aleja Lotników 32/46, PL-02-668 Warsaw, Poland}
\affiliation{International Research Centre MagTop, Institute of Physics. Polish Academy of Sciences, Aleja Lotnikow 32/46, PL-02-668  Warsaw, Poland}

\author{Jarosław~Wróbel}
\affiliation{Institute of Applied Physics, Military University of Technology, 2 Kaliskiego Str., PL-00-908 Warsaw, Poland}

\author{Piotr~Dziawa}
\affiliation{Institute of Physics, Polish Academy of Sciences, Aleja Lotników 32/46, PL-02-668 Warsaw, Poland}
\affiliation{International Research Centre MagTop, Institute of Physics. Polish Academy of Sciences, Aleja Lotnikow 32/46, PL-02-668 Warsaw, Poland}

\author{Tomasz~Wojtowicz}
\affiliation{International Research Centre MagTop, Institute of Physics. Polish Academy of Sciences, Aleja Lotnikow 32/46, PL-02-668 Warsaw, Poland}

\author{Tomasz~Story}
\affiliation{Institute of Physics, Polish Academy of Sciences, Aleja Lotników 32/46, PL-02-668 Warsaw, Poland}
\affiliation{International Research Centre MagTop, Institute of Physics. Polish Academy of Sciences, Aleja Lotnikow 32/46, PL-02-668 Warsaw, Poland}

\author{Jerzy~Wróbel}
\email{Corresponding author: wrobel@ifpan.edu.pl}
\affiliation{Institute of Physics, Polish Academy of Sciences, Aleja Lotników 32/46, PL-02-668 Warsaw, Poland}
\affiliation{Institute of Applied Physics, Military University of Technology, 2 Kaliskiego Str., PL-00-908 Warsaw, Poland}


\begin{abstract}
In this work, we used $20$~nm thick CdTe/SnTe/CdTe [001] quantum wells to make 6- and 8-terminal nano-structures with the etched cross-junctions of sub-micron width with walls directed along the [10], [01], and [11] surface crystallographic directions. We studied the low-temperature quantum magneto-transport to investigate the impact of lateral confinement on the states of topological carriers. Calculations showed that for narrow SnTe channels, almost flat bands with small energy dispersion are formed, and in the case of the [11] direction, the dispersionless states are strongly localized at the mesa edges. The measurements indicated that a current path associated with trivial states inside the quantum well was considerably narrowed due to disorder, leading to a significant reduction in channel conductivity. Such a high-resistance cross-junction has been used for measurements of non-linear transport in non-local configurations. The dependence of the differential resistance $R_\text{d}$ on the direct current $I_\text{DC}$ flowing through a selected pair of contacts was studied. For temperatures $T<1$~K, first an increase and then a decrease followed by a minimum of $R_\text{d}$ were observed. This is a characteristic $R_\text{d}(I_\text{DC})$ relationship that is often considered as the \textit{signature of the hydrodynamic flow} of a fermionic liquid in narrow quantum channels, which in the case of SnTe can be formed by topological states located entirely at the inner edges of a planar cross-junction.
\end{abstract}

\maketitle

% ARTICLE BODY
\section{Introduction}
Most of the phenomena associated with classical and quantum transport in semiconductor nanostructures can be described using the non-interacting electron model. Recently, however, more and more attention has been paid to the study of electron flow in quantum channels, the description of which goes beyond the Fermi gas model and requires taking into account the electron-electron interaction in scattering processes and the effects known from fluid hydrodynamics \cite{Narozhny2022}. The strength of the correlation interactions for a given material and a specific sample can be assessed using the $\alpha$ parameter, which is equal to the ratio of the average potential energy to the average kinetic energy of the electron liquid. As it is known, $\alpha = \gamma_\mathrm{d} r_s \propto m^* / \epsilon_s$, where $r_s$ is the average distance between charges expressed in the effective Bohr radius, $m^*$ is the effective mass and $\epsilon_s$ is the dielectric constant of the material. The numerical  coefficient $\gamma_\mathrm{d}$ depends on the dimensionality and increases for confined systems as the dimension is reduced.

When the parameter $\alpha \ll 1$, the influence of correlation effects is negligible and the model of non-interacting electrons is sufficient to describe transport phenomena. However, when $\alpha \gtrsim 1$, it may turn out that the dominant scattering mechanism is electron-electron collisions (\textit{e-e}). The relaxation time $\tau_{ee}$ for this type of process can be estimated from the formula \cite{Lucas2018}
\begin{equation}
	\tau_{ee}\sim \frac{1}{\alpha^2}\frac{\hslash E_\mathrm{F}}{(kT_e)^2},
	\label{eq:tau3}
\end{equation}
where $T_e$ is the temperature of electron gas and $E_\mathrm{F}$ is Fermi energy. As explained above, we expect very frequent (\textit{e-e}) collisions in samples with low electron concentration (large $r_s$), high effective mass and small dielectric constant.

It should be emphasized that during such scattering, the total momentum of the electron liquid \textit{does not change} due to the indistinguishability of quantum particles.  Processes in which momentum is conserved do not contribute to resistance (we ignore the so-called \textit{Umklapp} scattering), so the role of normal \textit{e-e} collisions is to direct electrons towards the edge of the sample. This leads to phenomena related to the hydrodynamic flow in conductive channels that are sufficiently narrow. One of them is the so-called \textit{Gurzhi effect} \cite{Gurzhi1968}, which leads to the \textit{decrease} of a metallic sample resistance with an increase of $T_e$ (in a certain temperature range). For semiconductor nanostructures, this effect was observed for the first time in Ref.~\cite{Jong1995}, where AlGaAs/GaAs heterojunction quantum wires with a high-mobility two-dimensional electron gas were studied. Later, the characteristic dependence of the sample's resistance on the temperature of electron gas $T_e$ was observed in microstructures made of graphene \cite{Bandurin2016}. Also, in the diluted 2D hole system in GaAs/AlGaAs quantum wells the viscous transport in the hydrodynamic regime was reported \cite{Kumar2023}.

In this paper, we describe a phenomenon similar to the Gurzhi effect, which was observed in submicron conductive channels made from a CdTe/SnTe/CdTe quantum well, grown in the [001] crystallographic direction. Tin telluride (SnTe) seems to be the worst candidate for observing hydrodynamic flows of fermionic liquids. It is characterized by $p$-type conductivity, which is caused by holes of high concentration ($p \approx \num{e20}\ \mathrm{cm}^{-3}$) and relatively low mobility ($\mu\approx \num{e3}\ \mathrm{cm}^2/\mathrm{Vs}$). As a narrow-gap semiconductor, SnTe has a small density of states effective mass  ($m^*_d \approx 0.1\, m_0$) for top valence band of light holes and moreover, its static dielectric constant is very large ($\epsilon_s\approx 1000$)~\cite{Ravich1970}.

Fortunately, tin telluride belongs to the new class of materials, so called \textit{topological crystalline insulators} (TCIs) in which gapless surface states, that are protected by crystal mirror symmetry, coexist with the non-topological (bulk) charge carriers \cite{Fu2011}. It is known that a band inversion in SnTe occurs at the four $L$ points of the Brillouin zone, therefore exactly four Dirac cones are expected on the boundary planes (001), (111), and (110) \cite{HsiehLinLiuEtAl2012}. The gapless surface states, with a linear energy  dispersion, were indeed observed on (001) and (111) surfaces of SnTe-class materials by angle-resolved photo-emission spectroscopy (ARPES) \cite{DziawaKowalskiDybkoEtAl2012, XuLiuAlidoustEtAl2012, TanakaRenSatoEtAl2012, Polley2014}. 

In this work we suggest that \textit{quasi}-1D quantization of topological states in planar sub-micron Hall bar structures can lead to the formation of edge channels with very low dispersion of energy ($m^* \rightarrow \infty$) in the cross-junction area, where the perpendicular parts of the structure intersect. The location of carriers at the edge significantly reduces the screening efficiency, which leads to an increase of the parameter $\alpha$, as compared to the bulk states. Additionally, a dominant role in electron-electron scattering plays the high-frequency dielectric constant $\epsilon_{\infty}$, which is much smaller than $\epsilon_s$ \cite{Ravich1970}. These are favorable conditions for the existence of the effects associated with the hydrodynamic flow of charges in SnTe nanostructures.
	
\section{Samples}
\label{sec:samples}
The SnTe quantum wells structures used in our studies were grown on a (001)-oriented CdTe/GaAs hybrid substrates. These hybrid substrates were grown in a separate MBE chamber dedicated to the II-VI materials and had a 4 micrometer thick CdTe layer grown on (100) oriented 2-inch epi-ready GaAs wafers with a 20 nm thick ZnTe buffer layer grown first. Before inserting the hybrid substrate into the IV-VI MBE growth chamber, the substrate was etched in HCl. The growth of a 20 nm thick SnTe QW layer followed the initial growth of an additional 50 nm thick CdTe buffer, also serving as the lower barrier, and the structure was finalized by the growth of 100 nm thick CdTe serving as the top barrier and cap. The structural quality of the quantum well structures was controlled in-situ by observing RHEED patterns and measuring specular spot intensity oscillations and ex-situ by high-resolution X-ray diffraction (HRXRD). Further details of the epitaxial growth of the SnTe wells are given in Refs. \cite{DybkoMazurWolkanowiczEtAl2018,Sulich2022}.

Smaller rectangular fragments of the size $5\ \mathrm{mm} \times 5\ \mathrm{mm}$ were cut out of the epitaxial wafers, the edges of which were parallel to the crystallographic directions [100] and [010]. These plates were then coated with a PMMA electron-sensitive layer. For the heating of the resist and for further thermal treatment, the \textit{low-temperature method}, previously developed for II-VI semiconductor quantum wells \cite{Majewicz2014}, was used. After e-beam lithography process, the samples were chemically etched to remove wafer material at exposed sites. The depth of etching was chosen so that the separating channels descended below the SnTe layer forming quantum wells (the so-called \textit{deep-etching}).

Such processes were used to make two types of measurement structures in the so-called Hall configuration, which differed in channel sizes and the number of voltage probes (see insets to Fig.~\ref{fig:G-sampB} and Fig.~\ref{fig:A-loc-i-b}). The first type of samples are 8-terminal  macro-structures with lithographic total length $L=1000\ \muup\mathrm{m}$ and width $W=100\ \muup\mathrm{m}$. They were used to characterize the SnTe quantum well. Most of the measurements presented in the paper were carried out on the second type of samples, namely on 6- and 8-terminal nano-structures with total channel length $L\approx 16\ \muup\mathrm{m}$ and lithographic widths of the order of $1\ \muup\mathrm{m}$. As a result of chemical etching, the physical widths of the nano-structures met the condition $W<1\ \muup\mathrm{m}$.

\begin{figure*}
	\begin{centering} 
		\includegraphics[scale=0.9]{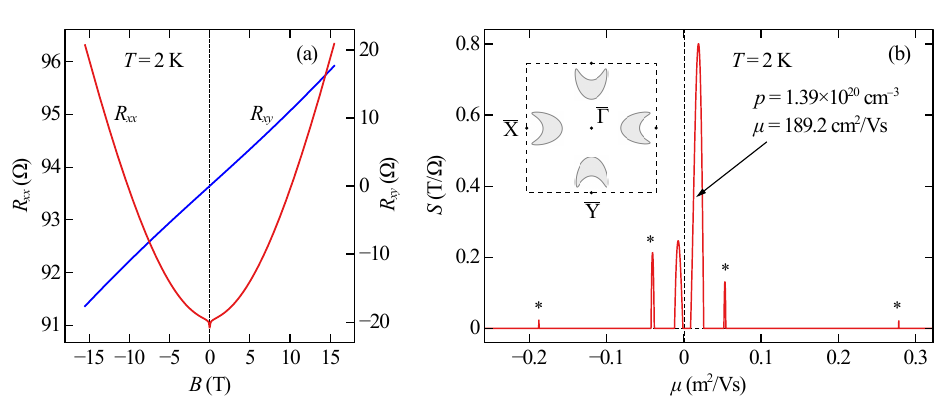} 
		\par\end{centering} 
	\caption{(a) Magnetoresistance $R_{xx}$ and Hall resistance $R_{xy}$ as a function of magnetic field $B$ for a macroscopic CdTe/SnTe(20nm)/CdTe quantum well sample at temperature $T=2\ \mathrm{K}$. (b) The mobility spectrum $S$ at $T=2\ \mathrm{K}$, concentration $p$ and mobility $\mu$ correspond to the dominant hole-like peak as indicated with arrow. Asterisks mark additional spectral lines associated with the presence of topological carriers. The inset shows (schematically) the shape of the constant energy curves for Fermi level $E_F \lesssim 26\ \mathrm{meV}$, dashed square depicts  the  2-dimensional Brillouin zone on crystallographic plane (100).\label{fig:Rij-dc-mode-spectrum}}	
\end{figure*}

The electrical parameters of the SnTe layer were investigated using conductivity and Hall effect measurements performed on 8-contact macroscopic samples. The measurements were made at a temperature $T = 2$~K using the constant-current (DC) method. Electrical connections were provided by indium contacts, soldered at a distance of about $2\ \mathrm{mm}$ from the center of the studied structure. The results are shown in Fig.~\ref{fig:Rij-dc-mode-spectrum}a, which demonstrates the dependence of the longitudinal resistance $R_{xx}$ and the Hall resistance $R_{xy} = R_{H}$ on the magnetic field~$B$.

Similarly to the SnTe quantum wells studied earlier \cite{DybkoMazurWolkanowiczEtAl2018}, in the entire range of magnetic fields we observe positive magnetoresistance and a clearly nonlinear relationship $R_H (B)$. In addition, for $B \le 2\ \mathrm{T}$ there is a narrow minimum of $R_{xx}$ associated with the effects of quantum interference of the wave function. In this case, it is the so-called \textit{weak anti-localization} (WAL), which has traditionally been considered as the evidence of topological states on the surface of SnTe. Here we will not analyze this effect, but we will show that the presence of 2-dimensional current carriers with an unusual dispersion relationship can be demonstrated on the basis of the analysis of classical transport.

For this purpose, the so-called Mobility Spectrum Analysis (MSA) was used, which is a standard method for studying classical transport in bulk samples and epitaxial structures \cite{McClure1956,BeckAnderson1987}. It consists in representing the components of the conductivity tensor, obtained from the experiment, in the form of integrals containing the products of Drude terms and the continuous function $S(\mu)$, called the \textit{mobility spectrum}. The observed spectral lines usually correspond to different types of carriers involved in the electrical conductivity of the sample. Figure \ref{fig:Rij-dc-mode-spectrum}(b) shows the function $S(\mu)$ derived from the data presented in Fig.~\ref{fig:Rij-dc-mode-spectrum}(a). By convention, the lines observed for $\mu<0$ correspond to the current carriers with the negative charge.

Therefore, the strongest spectral line visible in the figure corresponds to p-type carriers with concentration $p\approx \num{1.4e20}\ \mathrm{cm}^{-3}$ and mobility $\mu\approx 190\ \mathrm{cm}^2/\mathrm{Vs}$. These are undoubtedly native holes occupying the trivial states inside the quantum well, the obtained values are typical for CdTe/SnTe/CdTe epitaxial structures \cite{DybkoSzotSzczerbakowEtAl2017}. In addition, there is a weaker peak in the mobility spectrum, indicating the presence of electrons with much lower mobility. Their origin is not clear, but they are most likely related to the surface of the sample, which is covered with a relatively thick (100 nm) layer of CdTe. However, the presence of additional narrow spectral lines (marked with an asterisk), which occur almost symmetrically for much larger values of $|\mu|$, is noteworthy.  These peaks come from 2-dimensional topological carriers, located on the SnTe/CdTe interfaces and having a very unusual dispersion relationship (schematically shown in the inset).

The above conclusion is based on our previous work on SnTe epitaxial layers with thicknesses between 5 and 20 nm, also grown on the (001) plane \cite{Sniezek2023}. In the analyzed mobility spectra, pairs of narrow "satellite" peaks were observed, very similar to the spectral lines shown in Fig.\ref{fig:Rij-dc-mode-spectrum}(b). As shown by the simulation of classical transport, the shape of constant energy contours on the (001) surface results in electron-like and hole-like contributions to mobility spectrum. In other words, $p-$type and $n-$type peaks belong to the concave and convex parts of the deformed Dirac cones of a type shown in the inset. Moreover, this observation is valid in a relatively wide range of Fermi energies, when inner and outer Dirac cones develop \cite{LiuDuanFu2013}. Therefore, the results presented in Fig.~\ref{fig:Rij-dc-mode-spectrum}(b) strongly confirm the presence of characteristic satellite lines, previously observed for the SnTe/PbTe epilayers \cite{Sniezek2023}.

\section{1D quantization}
The unusual shape of Dirac cones and their location in the Brillouin zone have an impact not only on classical transport, but also on dimensional quantization of energy levels. In particular, the appearance of boundary states in thin layers of SnTe is expected \cite{Liu2014,Safaei2015} and the formation of 1-dimensional energy bands with zero dispersion on atomic steps occurring on the surface of bulk crystal is predicted \cite{Sessi2016,Rechcinski2018,Wagner2023}. In both cases, the thickness of films and the height of steps ranged from a few to a dozen of atomic layers, therefore the tight binding approximation (TBA) was mainly  used for calculations. In this work we are dealing with \textit{planar structures} of several hundred nanometers width, therefore the \textit{Kwant} package \cite{Groth2014} and $\boldsymbol{k}\cdot\boldsymbol{p}-$hamiltonian,  parameterized for (001) topological states \cite{LiuDuanFu2013}, were used for the analysis of dimensional quantization.

\begin{figure*}
	\begin{centering} 
		\includegraphics[scale=0.9]{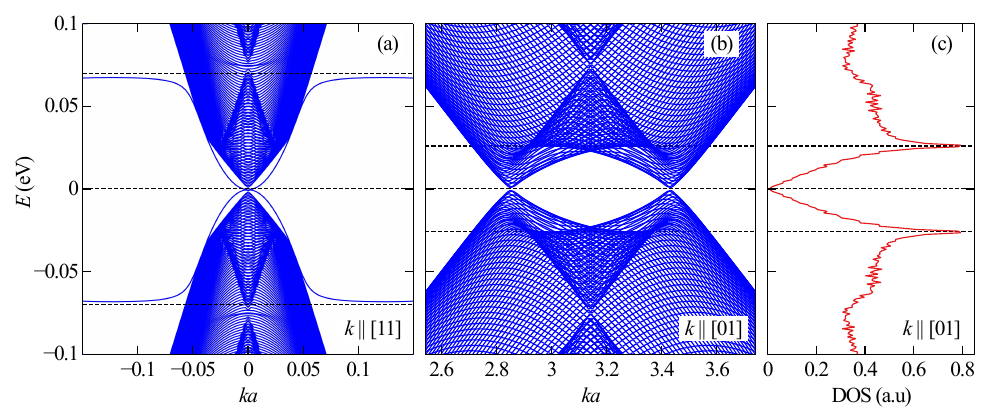} 
		\par\end{centering} 
	\caption{Energies $E$ of 1D levels and density of states (DOS) for quantum channels of width $W=Na$, where $a$ is the lattice step and $k$ is the quasi-momentum directed along the wire for two orientations. (a) $k \parallel [11]$,  $W=400\ \mathrm{nm}$ ($N=400$, $a=10\ \unit{\angstrom}$), values of $k$ close to the center of Brillouin zone (BZ). (b) $k \parallel [01]$,  $W=200\ \mathrm{nm}$ ($N=400$, $a=5\ \unit{\angstrom}$), values of $k$ close to the  boundary of BZ. (c) DOS calculated for $k \parallel [01]$, $W=400\ \mathrm{nm}$ ($N=800$, $a=5\ \unit{\angstrom}$).\label{fig:bands-dos-01}}	
\end{figure*}

The \textit{Kwant} program adopts the finite difference method for internal calculations, but in the case of linear dispersion the discretization of hamiltonian on 2-dimensional lattice leads to the so-called \textit{fermion doubling} \cite{Beenakker2023}. To avoid this notorious problem we used a method similar to the one described in \cite{Tworzydlo2008}, where finite differences are evaluated at points which are \textit{displaced} relative to the regular mesh nodes. Using this approach, we calculated the dispersion relations $E_i(k)$ for planar quantum wires of infinite length and width $W=Na$, where $a$ is the lattice parameter. In this case, $E_i$ is the energy of 1-dimensional level $i$, while $k$ is the quasi-momentum directed along the channel length. 

Figure~\ref{fig:bands-dos-01}(a) shows the dispersion relations, close to the $\bar{\Gamma}$ point, for topological states in a channel of width $W=400\ \mathrm{nm}$ oriented along $[11]$ surface crystallographic direction.  We see spatially quantized energy levels, the shape of which resembles projections of double Dirac cones on the channel axis. In addition, the results show the presence of two doubly degenerate \textit{flat bands}, whose energy $E \approx \pm 75\ \mathrm{meV}$ does not depend on the selected $a$ parameter. Moreover, the quantum states associated with this flat band are spatially confined \textit{to the edges} of a conductive channel. For other crystallographic directions, however, the energy pattern of the one-dimensional bands changes.

Figure~\ref{fig:bands-dos-01}(b) shows the dispersion relations for topological states in a channel of width $W=200\ \mathrm{nm}$ oriented along $[01]$ surface crystallographic direction. In this case, the quantized energy levels are pictured near the boundary of Brillouin zone. The results show again the presence of two almost \textit{flat bands}, this time with the camel-back shape, whose energy $E \approx \pm 26\ \mathrm{meV}$ practically does not depend on $k$. Apparently, these pseudo-bands are formed from many intersecting one-dimensional levels, grouped near the energy where two Dirac cones merge (so-called Lifshitz transition). Indeed, the density of states (DOS) shown in Fig.~\ref{fig:bands-dos-01}(c) reveals van Hove singularities for these energies.

The results shown in Fig.~\ref{fig:bands-dos-01} are presented in relatively narrow energy ranges because the curvature of the higher bands changed depending on the selected pitch of the discretization mesh. Fortunately, numerical tests showed that for small energies $E\lesssim 0.1\ \mathrm{eV}$ and wave vectors changes $|\delta k|\lesssim 0.15$~\AA$^{-1}$, the results depended solely on the total quantum channel width $W=Na$. The choice of the parameter $a$ did not have a significant effect on the band energies for values of the quasi-momentum $k$ near the Dirac points, provided that the lattice step did not exceed approx. $10$~\AA.

\section{Results}
Electrical measurements were carried out on two nano-structures (samples A and B) from wafers described in Sec.~\ref{sec:samples}. Sample A had an overall length of $L = 15.2\ \muup\mathrm{m}$, the distance between two pairs of voltage contacts was $L_1 = 5.6\ \muup\mathrm{m}$. Sample B had three pairs of contacts, spaced $L_2 = 3\ \muup\mathrm{m}$ and $L_3 = 5\ \muup\mathrm{m}$, the total length was $L = 16\ \muup\mathrm{m}$. The physical widths were $W \approx 0.95\ \muup\mathrm{m}$ and $W \approx 0.55\ \muup\mathrm{m}$ for samples A and B, respectively. For both structures, the conductive channels were directed along the surface crystallographic direction [10] (or [01]), which in turn was parallel to the sample edge.

\begin{figure} 
	\begin{center}
		\includegraphics[scale=0.9]{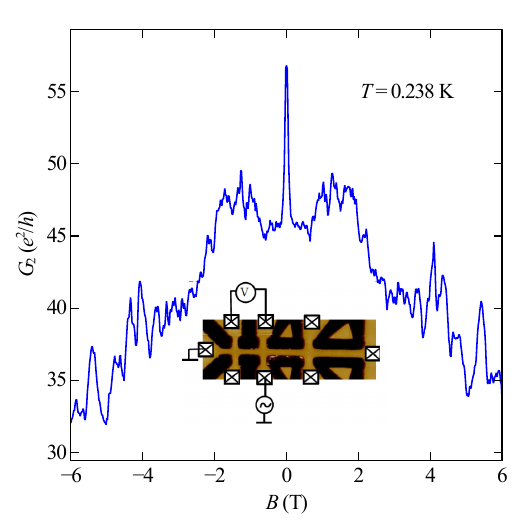}
	\end{center}
	\caption{Conductivity $G_2= 1/R_2$ as a function of magnetic field $B$ for temperature $T = 0.238\ \mathrm{K}$. $R_2$ is the differential resistance of \textit{sample B} measured on voltage probes separated by $L_2 = 3\ \muup\mathrm{m}$. The micro-graph of the structure and configuration of electrical contacts is shown in the inset.}
	\label{fig:G-sampB} 
\end{figure}

\begin{figure*}
	\begin{centering} 
		\includegraphics[scale=0.9]{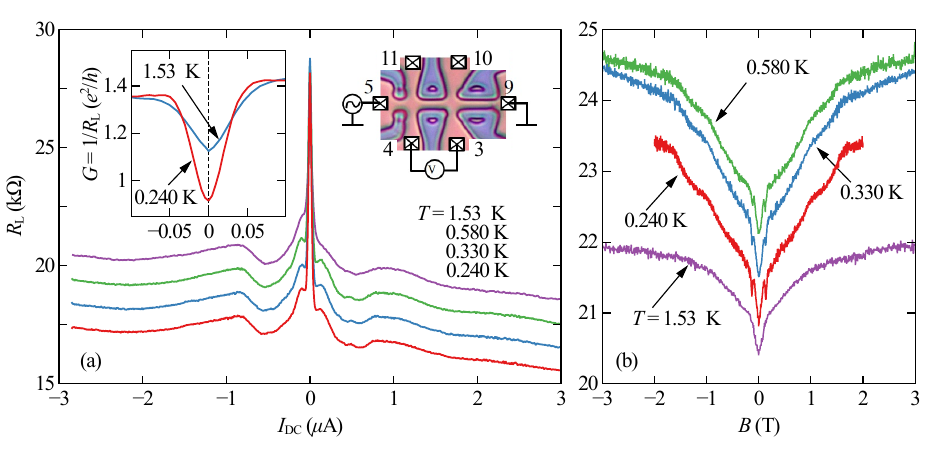} 
		\par\end{centering} 
	\caption{(a) Differential resistance $R_\mathrm{L}=R_{59,43}$ for sample A measured as a function of DC current $I_{DC}$ for different temperatures $T$. Configuration of voltage (3,4) and source-grain (5,9) probes is shown in the right inset. Curves for $T > 0.240\ \mathrm{K}$ are shifted upwards with a $1\ \mathrm{k}\Omegaup$ step. The left inset shows conductance $G=1/R$ in the range of small currents (without displacement along the $y$-axis). (b) Dependence of $R$ on the magnetic field $B$ and temperature for $I_{DC}\approx 0.025\ \muup\mathrm{A}$. The graphs for $T = 0.330\ \mathrm{K}$ and $T = 0.580\ \mathrm{K}$ are shifted upwards by $0.5\ \mathrm{k}\Omegaup$ and $1\ \mathrm{k}\Omegaup$, respectively. \label{fig:A-loc-i-b}}	
\end{figure*}

Quantum transport was studied in the temperature range from $0.240\ \mathrm{K}$ to $1.5 \mathrm{K}$ using the standard AC method. Macroscopic electrical contacts were made with silver paint, which were later replaced with indium soldered contacts. Despite this, some of the terminals (for both structures A and B) showed significant resistance at low temperatures, in the order of several M$\Omegaup$s. Therefore, not all voltage and current probe configurations were possible to examine. In some cases, the procedure of reheating and re-cooling to cryogenic temperatures has altered the properties of the samples themselves.

In general, it was found that the electrical properties of nano-structures differ significantly from those of macroscopic samples. For example, we expect a resistance of $R_1 \approx 600\ \Omegaup$ for sample A, after taking into account purely geometric factors. Similarly, for structure B, we expect resistance $R_3 \approx 500\ \Omegaup$ between contacts distant by $L_2 = 3\ \muup\mathrm{m}$. However, the measured resistances ranged from a few hundred ohms to several dozen k$\Omegaup$s, for different pairs of contacts. This may suggest that the physical widths of high-resistance  channels were much smaller than the geometrical width shown on micro-graphs. Moreover, the data obtained for such samples  suggested the presence of non-linear effects. We attribute these non-standard features of electrical transport in SnTe nano-structures to a greater role of topological carriers and changes in their band structure, caused by the confinement.

Figure \ref{fig:G-sampB} shows the conductivity of the nanostructure B, expressed in universal units $G_0=e^2/h$. For the tested pair of contacts (see inset), the resistance $R_2$ is about $500\ \Omegaup$, so it is close to expectations based on sample size. However, its dependence on the magnetic field is radically different from the data obtained for macroscopic sample. Firstly, $R_2$ grows much stronger in the magnetic field (for $B=6\ \mathrm{T}$ by as much as 60\%), and secondly, we observe reproducible oscillations, resembling universal conductivity fluctuations (UCFs), typical for disordered mesoscopic samples. However, their amplitude significantly exceeds the value of conductivity quantum $G_0$, which excludes their origin from the wave function interference. The same applies to the strong increase in conductivity, observed in a zero magnetic field. Its amplitude ($\Deltaup G \approx 10 G_0$) shows that it cannot be related to the weak anti-localization (WAL) effect observed in macroscopic sample (see Fig.~\ref{fig:Rij-dc-mode-spectrum}a), for which $\Deltaup G \approx  G_0$ at $T=0.4\ \mathrm{K}$.

A partial explanation of the observed phenomena can be provided in Fig.~\ref{fig:bands-dos-01}b, which shows one-dimensional topological states for a quantum channel with the same orientation and similar width as the studied sample.  We expect that as a function of the magnetic field, numerous energy states will be successively filled and emptied, as it happens with the change in Fermi energy $E_\mathrm{F}$. This can lead to fluctuations in conductivity, similar to the DOS oscillations seen in Fig~\ref{fig:bands-dos-01}c. Since very many states change their occupancy almost simultaneously, the amplitude of fluctuations can significantly exceed $e^2/h$, despite the presence of disorder. This conclusion was confirmed by quantum transport simulations carried out using the \textit{Kwant} package, however, the origin of the zero field anomaly (ZFA) remains unclear.

Transport measurements performed in other contact configurations confirmed that low-resistance and high-resistance channels coexist in studied structures. Below, we will focus on the results of nonlinear transport obtained for sample A. The central part of the structure showed much greater longitudinal resistance ($\approx 20\ \mathrm{k}\Omegaup$) than sample B, which nominally had a smaller conductive channel width. The study of nonlinear transport consisted in measuring the differential resistance $R_{ij,kl}=\mathrm{d}V_{kl}/\mathrm{d}I_{ij}$ as a function of constant (DC) voltage applied to current contacts. The results are presented in the form $R_{ij,kl}(I_\mathrm{DC})$, where $I_\mathrm{DC}$ is the direct current flowing through the source-drain terminals. All data shown in the figures are the mean values obtained from measurements for both directions of the $I_\mathrm{DC}$ sweep.

Fig.~\ref{fig:A-loc-i-b}a shows the longitudinal differential resistance $R_\mathrm{L}$ measured as a function of DC current, data are presented for four temperatures in the range from $0.240$ to $1.53\ \mathrm{K}$. For all curves the strong zero bias anomaly (ZBA) is observed, which is manifested by a sharp increase in the conductivity when $I_\mathrm{DC}$ changes from $0$ to approximately $0.05\ \muup\mathrm{A}$.  At lowest temperatures $\Delta G\approx 0.5\ e^2/h$ then it decreases, as shown in the inset to Fig.~\ref{fig:A-loc-i-b}a. Observed ZBA may be related to the presence of a tunnel barrier between the 2-dimensional carriers in the wide parts of the sample and the current-carrying narrow channels in the central part of the structure \cite{Gloos2009}.

Fig.~\ref{fig:A-loc-i-b}b shows the dependence of the longitudinal resistance on the magnetic field, measured slightly off-anomaly ($I_{DC}\approx 0.03\ \muup\mathrm{A}$). We observe a minimum for weak fields in the $B\approx\pm 0.1\ \mathrm{T}$ range, but the data differ quantitatively and qualitatively from those obtained for low-resistance channels. Compared to the data for sample B, shown in Fig.~\ref{fig:G-sampB}, the fluctuations of conductance, which were attributed to the population and depopulation of one-dimensional levels as a function of the magnetic field, are not observed. Nevertheless, at low temperatures, two or three shallow quasi-periodic oscillations are visible, the positions of which do not depend on temperature, only their amplitude decreases. Perhaps these are dimensional effects related to sample geometry.

The identification of the mechanism responsible for ZBA peak requires separate studies.  Therefore, below we will limit our analysis to the results obtained for $|I_{DC}|>0.05\ \muup\mathrm{A}$. As can be seen in Fig.~\ref{fig:A-loc-i-b}a, for both directions of the DC current, a small maximum of $R_\mathrm{L}$ is observed first, and then a wide longitudinal resistance minimum develops for $I_{DC}\approx 0.5\ \muup\mathrm{A}$. At low temperatures, the shape of this approximately symmetrical relationship does not change, only at $T = 1.53\ \mathrm{K}$ the smaller maxima are blurred and merge with the central ZBA peak. Very similar dependencies were observed in the measurements of non-local resistance and also in the non-linear transport data obtained for the Hall configuration, as discussed below.

\begin{figure} 
	\begin{center}
		\includegraphics[scale=0.9]{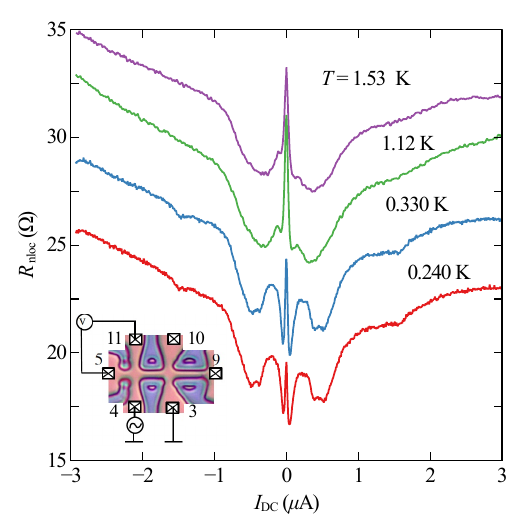}
	\end{center}
	\caption{Differential non-local resistance $R_\mathrm{nloc}=R_{34,511}$ for sample A (see inset) measured as a function of DC current $I_{DC}$ and temperature $T$. The graphs for $T > 0.240\ \mathrm{K}$ are shifted upwards with the $2.5\ \Omegaup$ step.} \label{fig:A-nloc}
\end{figure}

Figure~\ref{fig:A-nloc} shows the results of $R_\mathrm{nloc} = R_{34,511}$ measurements, made in a non-local configuration with the voltage probes located outside the current path, therefore $R_\mathrm{nloc} \ll R_\mathrm{L}$. Still, the zero bias anomaly was observed, however, with the height and width of the anomalous peak being much smaller this time. The difference may be related to the fact that now the different current probes were used. Nevertheless, the effects previously observed in the local configuration for $|I_{\mathrm{DC}}|>0.05\ \muup\mathrm{A}$, are seen even more clearly. The initial increase in resistance as a function of $I_\mathrm{DC}$ is much larger and the maximum is better pronounced. For higher lattice temperatures ($T = 1.12\ \mathrm{K}$ and $T = 1.53\ \mathrm{K}$) the maximum disappears, while a wide minimum is still observed.

\begin{figure} 
	\begin{center}
		\includegraphics[scale=0.9]{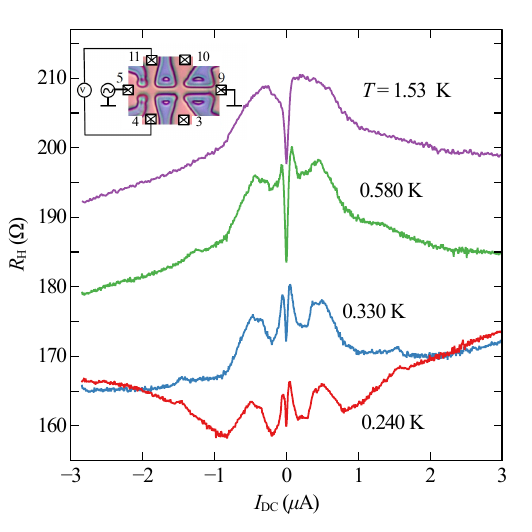}
	\end{center}
	\caption{Differential Hall resistance $R_\mathrm{H}=R_{59,411}$ for sample A (see inset) measured as a function of DC current $I_{DC}$ and temperature $T$. The graphs for $T > 0.240\ \mathrm{K}$ are shifted upwards with the $10\ \Omegaup$ step.} \label{fig:A-hall}
\end{figure}

\begin{figure*}
	\begin{centering} 
		\includegraphics[scale=0.9]{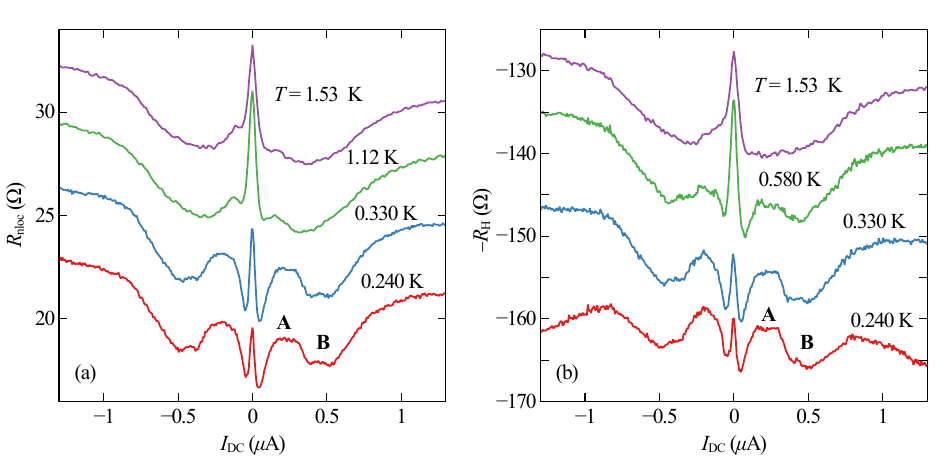} 
		\par\end{centering} 
	\caption{Comparison of the data shown in Figures~\ref{fig:A-nloc} and \ref{fig:A-hall} for $|I_{\mathrm{DC}}|<1.3\ \muup\mathrm{A}$. (a) Resistance measured in a non-local configuration. The curves for temperatures $T>0.240\ \mathrm{K}$ are shifted upwards with a $2.5\ \Omegaup$ step. (b) Resistance in the Hall configuration, shown \textit{with a minus sign}. The graphs for temperatures $T>0.240\ \mathrm{K}$ are shifted upwards with a $20\ \Omegaup$ pitch.} \label{fig:A-nloc-hall}
\end{figure*}

The results presented so far were obtained for voltage probes placed on the same side of the quantum channel. Figure~\ref{fig:A-hall} shows the differential resistance measured in the Hall configuration, in which contacts (5) and (9) again acted as current probes, while channels (4) and (11) on the opposite side of the sample were used to measure the voltage. At first glance, the data for Hall resistance differ from those obtained for voltage probes placed on the same edge. In reality, however, the dependence of $R_\mathrm{H}$ on the current $I_\mathrm{DC}$ flowing through the sample seems to be exactly in \textit{anti-phase} to the results obtained in non-local and local conductivity measurements with the ZBA peak pointing in the opposite direction. This is very clearly visible in Fig.~\ref{fig:A-nloc-hall}.

Panel (a) of Fig.~\ref{fig:A-nloc-hall} once again presents the results of differential resistance measurements in a non-local configuration, this time in a narrower range of currents $I_\mathrm{DC}$. On the right, panel (b) shows analogous curves for the Hall configuration, but the resistance values are taken there with \textit{a minus sign}. Measurements in both configurations were not performed simultaneously because different current probes were used and only one voltage probe (11) was common for both configurations. Still, there is a striking similarity between the two data sets. At low temperatures, the maximum (marked \textbf{A}) and then the wider minimum (\textbf{B}) are clearly visible, furthermore only feature \textbf{B} survives at higher temperatures. The same pattern is visible in Fig.~\ref{fig:A-loc-i-b} where maximum \textbf{A} is almost merged with the strong ZBA peak.

\section{Discussion}

Apart from the narrow peak associated with ZBA, the observed dependences of differential resistance on current $I_{\mathrm{DC}}$ are very similar to the analogous results obtained for AlGaAs/GaAs heterojunction quantum wires, reported in Ref.~\cite{Jong1995}. The main idea behind this pioneering work was that DC current increases only the temperature of the electron gas (but not of the crystal lattice) due to the weak coupling with phonons. A slight increase, then decrease and a minimum of resistance were observed, which was interpreted as an effect of the hydrodynamic flow of charges along a sufficiently narrow current path. Very similar relationships have recently been observed for microstructures made of graphene \cite{Bandurin2016}. It is the conductive channels made of this 2D material that have become the basic objects used to study hydrodynamic flow in a fermionic liquid \cite{Lucas2018}.

\begin{figure} 
	\begin{center}
		\includegraphics[scale=0.9]{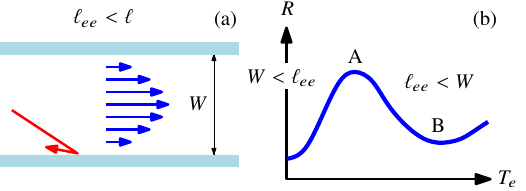}
	\end{center}
	\caption{(a) Distribution of the flow velocity of a fermionic liquid in a $W$ wide channel (blue arrows), in which there are  only (\textit{e-e}) collisions and scattering on the walls (red arrow), $\ell$ is mean free path, $\ell_{ee}$ is scattering length. (b) Resistance $R$ versus the temperature of electron liquid $T_e$ (schematically), the decrease of $R$ from maximum A to minimum B is referred as Gurzhi effect.}
	\label{fig:Hydro1} 
\end{figure}

Figure \ref{fig:Hydro1} shows the flow of an electron liquid in a narrow channel during which electron-electron scattering predominates. Since this scattering does not change the velocity (momentum is conserved), the only mechanism that changes the direction of the electron's motion is collisions with the channel walls. Therefore, the velocity distribution (blue arrows) is heterogeneous and the flow rate of the fermionic liquid decreases near the edges of the structure. The parameter that describes such a hydrodynamic velocity distribution is \textit{viscosity} $\eta$ (the so-called \textit{shear viscosity}), similar to the flow of a classic liquid that 'sticks' to the walls of a narrow capillary.

\begin{figure*}
	\begin{centering} 
		\includegraphics[scale=0.9]{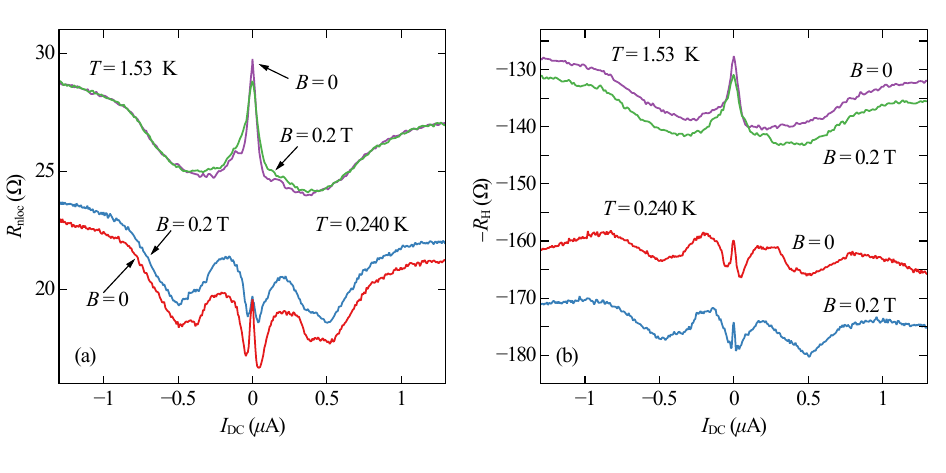} 
		\par\end{centering} 
	\caption {(a) Comparison of resistances measured in a non-local configuration for $B=0$ i $B=0.2\ \mathrm{T}$ at temperatures $0.240\ \mathrm{K}$ and $1.53\ \mathrm{K}$. Curve for $T=1.53\ \mathrm{K}$  is shifted up by $4\ \Omegaup$. (b) The same for Hall configuration (with minus sign). Data for $T=1.53\ \mathrm{K}$  are shifted up by $60\ \Omegaup$.} \label{fig:A-nloc-hall-B}
\end{figure*}

The viscosity $\eta$ in such a flow is proportional to the relaxation time $\tau_{ee}$ because the longer the free path $\ell_{ee} = v_\mathrm{F} \tau_{ee}$ the easier the electrons reach the walls. An example of a quantum liquid for which $\eta \propto \tau_{ee}$ is the liquid isotope of helium $^3\mathrm{He}$, whose atoms are fermions. The formula \ref{eq:tau3} shows that the viscosity of this liquefied gas should \textit{increase} at the lowest temperatures as $T^{-2}$, which is indeed observed in the experiments \cite{Huang2012}. An increase in viscosity down to the lowest temperatures is possible in Helium-3, because in a liquid we do not deal with scattering on lattice defects or scattering on phonons. On the other hand in solids, the interaction with the channel walls is less important in transport and the distribution of velocity across the direction of the current is constant. Then the effects related to the hydrodynamic flow of electrons are not expected.

Moreover, the $\alpha$ parameter is very small in metals  ($\alpha\approx r_s$) and scattering on ionized impurities dominates at low temperatures. Therefore, the most common relationship is $\ell < \ell_{ee} < W$, where $\ell$ is the average free path for scattering on dopants, and $W$ is the width of the conductive channel. Then the resistance of the sample is described by the mobility $\mu$ and the concentration $n$ of the carriers. However, when the condition $\ell_{ee} < W < \ell$ is met (see Fig.\ref{fig:Hydro1}), Gurzhi showed \cite{Gurzhi1968} that the resistance of the sample is determined by the viscosity parameter $\eta$ and the shape of the conductive channel. Since the viscosity of the liquid fermions decreases with increasing temperature, the resistance of the sample \textit{decreases} with an increase of $T_e$. For higher temperatures, the resistance can increase again due to scattering on phonons.

Additionally, for the lowest temperatures when the smallest length scale becomes the width of the channel $W< \ell_{ee} < \ell$, the resistance \textit{increases} with an increase of $T_e$. This is because the more frequent the electron-electron collisions (smaller $\tau_{ee}$), the more electron liquid is directed towards the edges of channel, where collisions with a change in momentum occur, contributing to $R$. This can be considered an electron analogue of the so-called, \textit{Knudsen effect}, known for the flow of gases in narrow capillaries of \cite{Narozhny2022}.

Going back to our data, we assume that at low temperatures, an increase of current $I_\mathrm{DC}$ leads to the heating of fermionic liquid, as in the other two-dimensional systems \cite{Jong1995,Bandurin2016}. Therefore, the observed maximum \textbf{A} and then the minimum \textbf{B} of differential resistance, may be a manifestation of the viscous Knudsen flow which evolves into the Gurzhi effect, as $T_e$ increases. However, the hydrodynamic flow, possibly responsible for the observed pattern, cannot occur for carriers residing in the quantum well. As estimated from the MSA calculations, mean free path of holes $\ell_h\approx 20\ \mathrm{nm} \ll W\approx 950\ \mathrm{nm}$, therefore we are dealing with the diffusive transport and for low-resistance channels we do not expect a noticeable effects related to viscous flow.

In the case of high-resistance channels, transport is most probably associated with the topological carriers and differential resistance is determined by the probabilities of transmission between distinct electrical terminals $\mathcal{T}_{ij}$. Obviously, each cross junction of the studied Hall structure contains edges which are oriented along the [10] and [01] surface crystallographic directions. However, in the central part, where perpendicular terminals intersect, boundary is parallel to the [11] axis. There, according to our calculations, states of flat-topological band form a very narrow edge channel of width $W_t \ll W$.

Furthermore, we expect that the mean free path of carriers located at the boundary $\ell_t > \ell$ due to the topological protection and therefore the condition $\ell_t > W_t$ can be easily met. Moreover, (\textit{e-e}) collisions are very frequent in the dispersionless band and scattering length  $\ell_{ee}$ may be small enough to satisfy the $\ell_{ee} < W_t$ relation.  This is also facilitated by a significant weakening of screening for states located on edges. Under these conditions, it is possible to observe the Gurzhi effect.

Therefore, we believe that the presented results show characteristic features of hydrodynamic flow. Nevertheless, additional commentary is required on the measurements carried out in the Hall configuration. In this case, the voltage contacts are on opposite sides of the current path and the measured asymmetry voltage is determined also by the transmission coefficients $\mathcal{T}_{ij}$ between the opposite edges of the sample. However, if due to disorder the transmission probability for closer channels is lower than for those that are more spatially separated, the measured voltage may be negative \cite{Datta1995}.

The important role of edge transport, which occurs in the area of cross junction, is confirmed by the results presented in Fig.~\ref{fig:A-nloc-hall-B}, where data shown in Fig.~\ref{fig:A-nloc-hall} are compared with the same results obtained for $B \neq 0$ at selected temperatures. For $T=0.240\ \mathrm{K}$, both $R_\mathrm{nloc}$ and $R_\mathrm{H}$ increase in the magnetic field $B=0.2\ \mathrm{T}$ by $\approx 1\ \Omegaup$ and $\approx 10\ \Omegaup$ , respectively, so in both cases by about 5\%. At the highest temperature tested, $T=1.53\ \mathrm{K}$, this increase is much smaller, which may indicate that the transmission coefficients $\mathcal{T}_{ij}(B)$ strongly depend on temperature. It is striking, however, that the curves move almost parallel with the magnetic field, which means that the \textit{asymmetry} $\pm I_\mathrm{DC}$ practically does not change in the studied field range. This means that we do not observe effects associated with magnetic focusing, suggesting that at least part of the topological carrier transport in the cross-junction area is mediated by edge states. The only change in shape that we observe for $B=0.2\ \mathrm{T}$ is a decrease in the height of ZBA peak at low temperatures.

\section{Summary}

We studied $20$~nm thick CdTe/SnTe/CdTe [001] quantum wells (QWs) in which the  existence of gapless surface states, together with the non-topological charge carriers (holes), was confirmed by the Mobility Spectrum Analysis (MSA). From such wafers we made 6- and 8-terminal nano-structures with etched channels of sub-micron width, formed as two or three 4-terminal cross-junctions connected in series along the [10] surface crystallographic direction. We studied the low-temperature quantum magneto-transport, motivated by the effect of spatial confinement on SnTe surface states. Our calculation showed that on the grid of intersecting one-dimensional levels, almost flat bands with small dispersion are formed, for which the density of states has a maximum. In the case of [11] direction, the dispersionless states are \textit{strongly localized at the edges} of a channel.

The results of the measurements showed that the transport properties of nano-structures differ significantly from macroscopic samples. In particular, the resistance of some of the etched channels was much greater ($R>10$~k$\Omegaup$) than the geometric size implied. Apparently, a current path associated with the holes from the quantum well was considerably narrowed due to disorder, and overall conductance was reduced. In such channels, unusual dependencies of local and non-local differential resistance on source-drain voltage, magnetic field, and temperature were observed. The obtained data suggest that in the studied structure, the current flow involves dispersionless carriers located on the lateral edges of the cross-junction area, oriented in the [11] surface crystallographic direction. With this assumption, the minima of local and non-local resistances, observed as a function of DC current flowing through the sample, were interpreted as a manifestation of the Gurzhi effect, related to the hydrodynamic flow of the fermionic liquid.

To summarize, in narrow SnTe channels, lithographically made from quantum wells, the contribution of topological carriers to transport phenomena increased significantly. This allowed for at least partial coping with the problem typical of SnTe, which is the high concentration of holes occupying the trivial quantum states and dominating the current flow. At the same time, a significant reduction in the width of the quasi-one-dimensional channels below $1\ \mu$m is not required, which makes it possible to produce planar structures with more complex geometries. In particular, cross-junctions, similar to those studied in this work, allow access to short quantum channels oriented towards [11] direction, with potentially interesting properties \cite{Kawala2024}. Our calculations suggest that the electric current can then partially flow along edge states occupying the flat band. For this type of sample, we observed signs of viscous flow of topological carriers and found the presence of the so-called Zero Bias Anomaly (ZBA).

\begin{acknowledgments}	
	
This research was partially supported by the ``MagTop'' project (FENG.02.01-IP.05-0028/23) carried out within the "International Research Agendas" program of the Foundation for Polish Science co-financed by the European Union under the European Funds for Smart Economy 2021–2027 (FENG). Jarosław Wróbel and Jerzy Wróbel acknowledge the additional support of Polish Ministry of Science and Higher Education through the “Polish Metrology” program, project No. PM/SP/0036/2021, budget: 880 000,00 PLN.
\end{acknowledgments}

% bibliography
%\bibliographystyle{apsrev4-2}
%\bibliography{./bib/PhD_Sniezek_final_nano_2024}
% from bbl
%apsrev4-2.bst 2019-01-14 (MD) hand-edited version of apsrev4-1.bst
%Control: key (0)
%Control: author (72) initials jnrlst
%Control: editor formatted (1) identically to author
%Control: production of article title (-1) disabled
%Control: page (0) single
%Control: year (1) truncated
%Control: production of eprint (0) enabled
%
%	
\end{document}